\documentclass[%
reprint,
 amsmath,amssymb,
 aps,
]{revtex4-2}
\usepackage{amsmath}
\usepackage{amssymb}
\usepackage{physics}
\usepackage{graphicx}
\usepackage{dcolumn}
\usepackage{bm}

\usepackage{hyperref}
\usepackage{xcolor}

\hypersetup{
    colorlinks,
    citecolor=blue,
    filecolor=blue,
    linkcolor=blue,
    urlcolor=black
}

\newcommand{\red}[1]{\textcolor{black}{#1}}

\begin{document}

\preprint{APS/123-QED}

\title{Mesoscopic theory of the Josephson junction}

\author{Thomas J. Maldonado}
\email{maldonado@princeton.edu}
\affiliation{%
Department of Electrical and Computer Engineering, Princeton University, Princeton, NJ 08544, USA}
\author{Hakan E. Türeci}%
\affiliation{%
Department of Electrical and Computer Engineering, Princeton University, Princeton, NJ 08544, USA}
\author{Alejandro W. Rodriguez}%
\affiliation{%
Department of Electrical and Computer Engineering, Princeton University, Princeton, NJ 08544, USA}

\begin{abstract}
We derive a mesoscopic theory of the Josephson junction from non-relativistic scalar electrodynamics. Our theory reproduces the Josephson relations with the canonical current phase relation acquiring a weak second harmonic term, and it improves the standard lumped-element descriptions employed in circuit quantum electrodynamics by providing spatial resolution of the superconducting order parameter and electromagnetic field. By providing an \emph{ab initio} derivation of the charge qubit Hamiltonian that relates traditionally free qubit parameters to geometric and material properties, we progress toward the quantum engineering of superconducting circuits at the subnanometer scale.
\end{abstract}

\maketitle
\begin{figure}[ht!]
\includegraphics{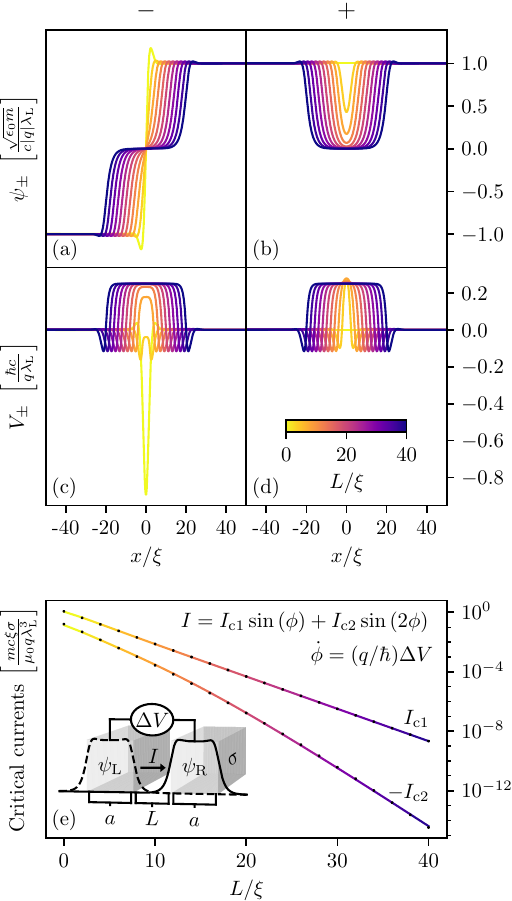}
\caption{\label{fig} Derivation of the Josephson relations begins by solving the EOM [Eqs.~\eqref{eq:EOM_simple}] for the stationary parity states (a) \(\psi_-\) and (b) \(\psi_+\), along with their electric potentials (c) \(V_-\) and (d) \(V_+\) in the presence of a background charge density [Eq.~\eqref{eq:background}] representing the junction geometry in (e) with \(a\gg \xi\) and \(\sigma \gg \xi^2\). In the limit of small charge imbalance, the wave function may be expressed in the left/right basis [Eq.~\eqref{eq:LRbasis}] shown in (e) as \(\psi \approx (\psi_\text{L} + e^{i\phi}\psi_{\text{R}})/\sqrt{2}\). The EOM then relate the current \(I\), phase difference \(\phi\), and voltage drop \(\Delta V\) across the junction via the Josephson relations in (e) with the critical currents \(I_{\text{c}1,2}\) and the positions of the two electrodes measuring \(\Delta V\) uniquely determined by \(\psi_\pm\) and \(V_\pm\). The former is plotted (solid lines) in (e), and the latter reduces to the two island centers for \(a\gg\xi\), as illustrated in (e). Extrapolation of the critical currents to larger junction sizes is made possible with the quadratic fits plotted (dotted lines) in (e): \(\abs{I_{\text{c}{1,2}}}\approx[mc\xi\sigma/(\mu_0\abs{q}\lambda_\text{L}^3)]\exp\{a_{1,2}(L/\xi)^2 + b_{1,2}(L/\xi)+c_{1,2}\}\) with \(a_1\approx0.000\), \(b_1\approx-0.507\), \(c_1\approx0.155\) and \(a_2\approx-0.005\), \(b_2\approx-0.590\), \(c_2 \approx -1.819\).}
\end{figure}

The Josephson junction (JJ), a circuit element comprising two superconductors separated by a thin insulating region, is a key element in modern quantum technologies, including superconducting charge qubits for quantum computing~\cite{kjaergaard2020superconducting}, Josephson parametric amplifiers (JPAs) for quantum readout~\cite{aumentado2020superconducting}, and superconducting quantum interference devices (SQUIDs) for magnetometry~\cite{clarke2006squid}. The advantages of these technologies over traditional electronic components stem from the JJ’s ability to exhibit nonlinear dynamics driven by quantum tunneling of superconducting charge carriers on a macroscopic scale \cite{josephson1962possible}. Noteworthy manifestations of these nonlinearities include the DC and AC Josephson effects~\cite{likharev2022dynamics}, both of which arise from the celebrated Josephson relations governing the flow of charge through the junction \cite{josephson2}.

The literature is replete with microscopic derivations of the Josephson relations. In their simplest form, these theories employ the well-known tight binding Hamiltonian with a phenomenological hopping parameter that determines the critical current~\cite{PhysRevB.3.769}. Various extensions have been considered via the electromagnetic response functions of the Bardeen-Cooper-Schrieffer (BCS) theory~\cite{bardeen1957theory, PhysRev.111.412}, but the complexity introduced by vast microscopic degrees of freedom has long posed a challenge to the analysis of generalized time dynamics~\cite{likharev2022dynamics}. As a consequence, further progress has historically relied on the introduction of additional phenomenological parameters (e.g., the junction's intrinsic capacitance), without insight into their dependence on the JJ's geometric form or photonic environment. We note that while more \emph{ab initio} methods are possible, they often rely on the calculation of a tunneling matrix between an impractically large number of entries that nonetheless assume a phenomenological barrier strength. 

Macroscopic derivations of the Josephson relations offer a simpler alternative: in lieu of the many-body BCS ground state used in microscopic derivations, these theories focus on the superconducting order parameter central to the Ginzburg-Landau (GL) theory~\cite{GL, citro2024josephson}. The GL theory's ability to model steady-state superconducting phenomena in inhomogeneous domains holds promise for a geometrically informed description of the JJ, but its canonical form is not a dynamical one~\cite{oripov2020time}. Derivations employing a two-component order parameter (one per side of the junction) have accordingly turned to Schrödinger's equation for time dynamics. While simple, this framework still requires a phenomenological Hamiltonian coupling the two components~\cite{feynman}, and the spatial partitioning of the order parameter used therein forsakes any description of its (or the electromagnetic field's) spatial distribution.

Some progress has been made to remedy the aforementioned shortcomings in related systems, such as trapped Bose-Einstein condensates via the Gross-Pitaevskii (GP) equation \cite{ostrovskaya2000coupled, salasnich1999bose, saeed, PhysRevA.59.620}, but the inability of both the BCS and GL theories to effectively model inhomogeneous electrodynamic phenomena in a unified domain poses a challenge in superconducting systems. With the rise of increasingly complex superconducting circuitry in quantum devices, a mesoscopic description of the JJ that captures its spatial characteristics and interactions with electromagnetic surroundings is long overdue. \red{Here, ``mesoscopic" refers not to the physical dimensions of the JJ~\cite{barone1982physics}, but rather to a mean-field treatment of the order parameter that  (without resolving the microscopic BCS degrees of freedom) incorporates its nonlocal spatial variations across superconducting and non-superconducting domains (a feature typically absent in macroscopic GL-based models).}

In this work, we present a mesoscopic theory of the JJ by applying the theory of non-relativistic scalar electrodynamics to the superconducting order parameter at low temperatures~\cite{PhysRevB.110.014508, PhysRevA.107.053704}. In doing so, we provide an \emph{ab initio} derivation of the Josephson relations that relates traditionally free parameters, such as the JJ's critical current and intrinsic capacitance, to both geometric and material properties of the JJ. Focusing on a simple, instructive geometry of coplanar superconducting islands separated by vacuum, we predict a critical current that decays exponentially with the separation length, along with a correction to the canonical current phase relation (CPR) in the form of a weak second harmonic term~\cite{willsch2024observation}. Finally, by deriving the charge qubit Hamiltonian from first principles, we connect to the framework of circuit quantum electrodynamics (cQED) \cite{blais2021circuit} in a way that relates crucial quantum informational parameters like the qubit anharmonicity to geometric degrees of freedom. This approach provides a foundation for the quantum engineering of superconducting circuits in more general geometries.

Throughout the text, we employ Einstein notation with the Minkowski metric \(\eta_{\mu\nu} = \text{diag}(+,-,-,-)_{\mu\nu}\). In accordance with convention, we use the Greek alphabet to index both temporal and spatial components and the Latin alphabet for purely spatial components. Our model for the superconducting order parameter \(\psi\) and electromagnetic four-potential \(A_\mu\) is defined by the Lagrangian for non-relativistic scalar electrodynamics under minimal light-matter coupling
\begin{multline}\label{eq:lagrangian}
    \mathcal{L} = \frac{i\hbar c}{2}\left(\psi^*\left(D_0\psi\right) - \psi \left(D_0\psi\right)^*\right) - \frac{\hbar^2}{2m}\left(D_i\psi\right)^*D_i\psi \\-\frac{1}{4\mu_0}F^{\mu\nu}F_{\mu\nu}-A^\mu j_\mu,
\end{multline}
where \(D_\mu = \partial_\mu + (iq/\hbar)A_\mu\) is the gauge covariant derivative, \(F_{\mu\nu} = \partial_\mu A_\nu - \partial_\nu A_\mu\) the electromagnetic tensor, \(j_\mu\) the four-current generated by non-superconducting (normal) charge carriers, and \(q\) and \(m\) the charge and mass of the superconducting bosons, respectively. The equations of motion (EOM) arising from this Lagrangian couple Schrödinger's equation for the order parameter to Maxwell's equations for the four-potential
\begin{subequations} 
\label{eq:maxwell_schrodinger}
    \begin{gather}
        i\hbar c D_0 \psi = -\frac{\hbar^2}{2m}D_i D_i \psi\\
        \partial_\mu F^{\mu\nu} = \mu_0 \left(\mathcal{J}^\nu + j^\nu\right),
    \end{gather}
\end{subequations}
where \(\mathcal{J}_\mu\) is the supercurrent, whose components are given by \(\mathcal{J}_0 = cq\abs{\psi}^2\) and \(\mathcal{J}_i = (i\hbar q/(2m))(\psi^*(D_i\psi)-\psi (D_i\psi)^*)\). 

We proceed by imposing two symmetries. First, we assume translational symmetry in two of the spatial dimensions with no normal current or vector potential (\(j_i = A_i = 0\)). Under this assumption, the EOM reduce to
\begin{subequations}\label{eq:EOM_simple}
    \begin{gather}
        i\hbar\dot{\psi} = \left(-\frac{\hbar^2}{2m}\laplacian + qV \right)\psi\label{eq:schrodinger}\\
        -\laplacian V = \frac{1}{\epsilon_0}\left(q \abs{\psi}^2 + \rho\right),\label{eq:gauss}
    \end{gather}
\end{subequations}
where Gauss's law [Eq.~\eqref{eq:gauss}] is the only one of Maxwell's equations not automatically satisfied, and we have used the shorthand \(V\equiv cA_0\) and \(\rho \equiv j_0/c\) for the electric potential and background ionic charge density, respectively. Second, we assume the background exhibits reflection symmetry in the remaining spatial dimension \(\rho(-x) = \rho(x)\), which allows us to identify time-independent solutions to Eqs.~\eqref{eq:EOM_simple} comprising stationary parity states \(\psi_\pm(-x) = \pm\psi_\pm(x) \in \mathbb{R}\) and their self-consistent electric potentials \(V_\pm(-x) = V_\pm(x)\). 

While the following analysis only requires that \(\rho\) respect these two symmetries, we focus on piece-wise constant (jellium) backgrounds taking values \(\rho = -qn_\text{s}\) inside and \(\rho= 0\) outside the material with \(n_\text{s}\) the bulk superconducting number density. In this case, both \(\psi_\pm\) and \(V_\pm\) vary on a length scale given by the healing length
\begin{equation}
    \xi \equiv \sqrt{\frac{\lambda_\text{L}\lambda_\text{C}}{4\pi}}
\end{equation}
with \(\lambda_\text{L}\equiv \sqrt{m/(\mu_0 n_\text{s}q^2)}\) the London penetration depth and \(\lambda_\text{C} \equiv h/(mc)\) the Compton wavelength of the bosons~\cite{PhysRevB.110.014508}. 
See Figs.~\ref{fig}(a)-(d) for examples of \(\psi_\pm\) and \(V_\pm\) solved numerically in the presence of two superconducting islands of length \(a\gg \xi\) separated by a vacuum region of length \(L\), as summarized by the shaded background in Fig.~\ref{fig}(e)
\begin{equation}\label{eq:background}
    \rho = -qn_\text{s}\left(\Theta(\abs{x-L/2})-\Theta\left(\abs{x-\left(a+L/2\right)}\right)\right)
\end{equation}
with \(\Theta\) the Heaviside step function. 

We begin our dynamical analysis by assuming the wave function may be represented as a linear combination of the stationary parity states. This two-mode approximation is summarized by the ansatz
\begin{equation} \label{eq:ansatz}
    \psi = \sum_\pm \alpha_\pm(t)\psi_\pm(x) = \sum_{\text{L},\text{R}}\alpha_{\text{L},\text{R}}(t)\psi_{\text{L},\text{R}}(x),
\end{equation}
where in the second equality of Eq.~\eqref{eq:ansatz}, we have introduced a basis of wave functions localized to the left and right of the axis of symmetry, which (with the convention \(\psi_\pm(\infty) > 0\)) are related to the stationary parity states by a 2D Hadamard transformation
\begin{subequations}
    \begin{align}
    \begin{pmatrix}
        \psi_\text{R}\\\psi_\text{L}
    \end{pmatrix}
    &\equiv
    \frac{1}{\sqrt{2}}
    \begin{pmatrix}1&1\\1&-1\end{pmatrix}
    \begin{pmatrix}
        \psi_+\\\psi_-
    \end{pmatrix}\label{eq:LRbasis}\\
    \begin{pmatrix}
        \alpha_\text{R}\\\alpha_\text{L}
    \end{pmatrix}&\equiv
    \frac{1}{\sqrt{2}}
    \begin{pmatrix}1&1\\1&-1\end{pmatrix}
    \begin{pmatrix}
        \alpha_+\\\alpha_-
    \end{pmatrix}.
    \end{align}
\end{subequations}
An example of the left/right basis functions solved for a finite island geometry \red{is plotted in dashed/solid black curves in  the inset of} Fig.~\ref{fig}(e). Plugging the ansatz [Eq.~\eqref{eq:ansatz}] into Schrödinger's equation [Eq.~\eqref{eq:schrodinger}] gives
\begin{subequations}\label{eq:2mode_parity}
    \begin{gather}
        i\hbar
        \begin{pmatrix}
            \dot{\alpha}_+\\\dot{\alpha}_-
        \end{pmatrix}
        =\Delta
        \begin{pmatrix}
            \alpha_+\\
            \alpha_-
        \end{pmatrix}\label{eq:2mode_parity_just_dynamics}
        \\
        \Delta_{ij} = \frac{q}{\mathcal{N}}\int_{-\infty}^\infty \left(V-V_j\right)\psi_i\psi_j dx\label{eq:delta_defs}
    \end{gather}
\end{subequations}
with \(\mathcal{N} \equiv \int_{-\infty}^\infty \abs{\psi_\pm}^2dx = N/\sigma\) the 1D normalization constant, \(N\) the total number of bosons and \(\sigma\) the cross-sectional area of the system depicted in \red{the inset of} Fig.~\ref{fig}(e). \red{The indices \(i,j\) run over the signs \(\pm\).} Equations~\eqref{eq:2mode_parity} provide a full description of the dynamics in the two-mode approximation, which may equivalently be expressed by the left/right coefficients in the rotating frame \(\tilde{\alpha}_{\text{L},\text{R}} \equiv \alpha_{\text{L},\text{R}} \exp{i\int _{-\infty}^t (\Delta_{++}+\Delta_{--})dt/(2\hbar)}\)
\begin{subequations}\label{eq:2mode}
    \begin{gather}
    i\hbar
        \begin{pmatrix}
            \dot{\tilde{\alpha}}_\text{R}\\
            \dot{\tilde{\alpha}}_\text{L}\\
        \end{pmatrix}
         = \begin{pmatrix}
             U &K\\K&-U
         \end{pmatrix}
         \begin{pmatrix}
             \tilde{\alpha}_\text{R}\\\tilde{\alpha}_\text{L}
         \end{pmatrix}\label{eq:2mode_just_dynamics}
         \\
         K \equiv \frac{\Delta_{++} - \Delta_{--}}{2}\label{eq:first_K}\\
        U \equiv \Delta_{\pm\mp}\label{eq:first_U},
    \end{gather}
\end{subequations}
as derived in Appendix~\ref{sec:2mode}.

We now evaluate the matrix elements \(K\) and \(U\). This requires first solving Gauss's law for the electric potential [Appendix~\ref{sec:potential}]
\begin{multline}\label{eq:potential}
    V= \frac{q}{\epsilon_0}\left(\abs{\tilde{\alpha}_\text{R}}^2-\abs{\tilde{\alpha}_\text{L}}^2\right)\int_0^x \int_{x'}^{\infty} \prod_\pm\psi_\pm(x'')dx''dx'\\+V'_\text{ext}(t)x+\sum_\pm\left(\frac{1}{2}\pm\Re{\tilde{\alpha}_\text{L}^*\tilde{\alpha}_\text{R}}\right)V_\pm,
\end{multline}
with \(V'_\text{ext} \equiv \lim_{\abs{x}\rightarrow \pm\infty}\partial_x V\) the negated external electric field, and then plugging in the result to yield the  matrix elements [Appendix~\ref{sec:matrix_elements}]
\begin{subequations}\label{eq:K}
    \begin{gather}
        K = K_1 +2\Re{\tilde{\alpha}_\text{L}^* \tilde{\alpha}_\text{R}}K_2\\
        K_1 = \frac{q}{4\mathcal{N}}\int_{-\infty}^\infty\left(\abs{\psi_-}^2 + \abs{\psi_+}^2\right)\left(V_--V_+\right) dx
        \\
        K_2 = \frac{q}{4\mathcal{N}}\int_{-\infty}^\infty\left(\abs{\psi_-}^2 - \abs{\psi_+}^2\right)\left(V_--V_+\right) dx.
    \end{gather}
\end{subequations}
and 
\begin{subequations}\label{eq:U}
    \begin{gather}
        U = \frac{qV'_\text{ext}d}{2} + \frac{q^2N}{4C_{\text{J}}}\left(\abs{\tilde{\alpha}_\text{R}}^2 - \abs{\tilde{\alpha}_\text{L}}^2\right )\\
        C_\text{J} = \frac{N^2\epsilon_0}{8\sigma}\left[\int_0^\infty\int_0^x\int_{x'}^\infty\prod_\pm\psi_\pm(x)\psi_\pm(x'')dx''dx'dx\right]^{-1}\label{eq:C_J}\\
        d \equiv \frac{4}{\mathcal{N}}\int_0^{\infty}x \psi_+\psi_-dx\label{eq:d}
    \end{gather}
\end{subequations}
with \(C_\text{J}\) the capacitance and \(d\) the \red{effective dipole separation. We postpone justification for these namings until later in the text but nonetheless note that \(d\) is proportional to the transition dipole moment between the stationary parity states and is thus the expected figure of merit to couple to the external electric field.}

In order to derive an effective Hamiltonian for these dynamics, we introduce the Madelung representation for the left/right coefficients, the phase difference across the junction, and the fractional population imbalance 
\begin{subequations}
    \begin{gather}
        \tilde{\alpha}_{\text{L},\text{R}}\equiv\sqrt{n_{\text{L},\text{R}}}e^{i\phi_{\text{L},\text{R}}}\\
        \phi \equiv \phi_\text{R}-\phi_\text{L}\\
        n \equiv n_\text{L}-n_\text{R},
    \end{gather}
\end{subequations}
in terms of which Eq.~\eqref{eq:2mode_just_dynamics} reads
\begin{subequations}\label{eq:n_phi_EOM}
    \begin{gather}
        \hbar\dot{n} = {2K}\sqrt{1-n^2}\sin\phi\\
        \hbar\dot{\phi} = -2U - \frac{2Kn\cos\phi}{\sqrt{1-n^2}}
    \end{gather}
\end{subequations}
with the matrix elements given by
\begin{subequations}
    \begin{gather}
        K = K_1 + K_2\sqrt{1-n^2}\cos\phi 
        \\
        U = \frac{qV'_\text{ext}d}{2}-\frac{q^2Nn}{4C_\text{J}}.
    \end{gather}
\end{subequations}
After the substitution \(\phi = 2\pi\Phi/\Phi_0\) and \(n = 2Q/Q_0\), with \(\Phi_0 = h/q\) the flux quantum and \(Q_0 = qN\) the total superconducting charge, these EOM are simply Hamilton's equations for the electric charge \(Q\) on the left (\(-Q\) on the right) and magnetic flux \(\Phi\) following from the Hamiltonian [Appendix \ref{sec:hamiltonian}]
\begin{multline} \label{eq:hamiltonian}
        H = -dV'_\text{ext}Q  + \frac{Q^2}{2C_\text{J}}\\- E_{\text{J}1}\sqrt{1-\left({2Q}/{Q_0}\right)^2}\cos\left(2\pi\Phi/\Phi_0\right)\\- 2E_{\text{J}2}\left({1-\left({2Q}/{Q_0}\right)^2}\right)\cos^2\left(2\pi\Phi/\Phi_0\right)
\end{multline}    
with the Poisson bracket \(\{\Phi,Q\} = 1\) and the Josephson energies 
\begin{subequations}~\label{eq:energies}
    \begin{gather}
        E_{\text{J}1} \equiv -NK_{1} = \frac{E_- - E_+}{2}\label{eq:EJ1}\\
        E_{\text{J}2} \equiv -\frac{NK_2}{4} = -\frac{\epsilon_0\sigma}{16}\int_{-\infty}^\infty \abs{\grad\left(V_--V_+\right)}^2dx.
    \end{gather}
\end{subequations}
where
\begin{equation}\label{eq:parity_energies}
    E_\pm = \sigma \int_{-\infty}^\infty \left(\frac{\epsilon_0}{2}\abs{\grad V_\pm}^2 + \frac{\hbar^2}{2m}\abs{\grad\psi_\pm}^2\right)dx
\end{equation}
are the energies of the stationary parity states, studied here~\cite{PhysRevB.110.014508} through the lens of electrohydrostatics. \red{That \(C_\text{J}\) represents a capacitance is now manifest by the second term in the Hamiltonian, which takes the canonical form \(Q^2/(2C_\text{J})\) of a capacitive chraging energy, and that \(d\) represents an effective dipole separation is now manifest by the first term in the Hamiltonian, which takes the canonical form \(pV'_\text{ext}\) of an electric dipole potential energy with the effective dipole moment \(p = -Qd\)}. That the coupling \(K_1\) (and therefore \(E_{\text{J}1}\)) scales with the energy splitting of the stationary parity states has been recognized in recent analyses of trapped BECs~\cite{PhysRevA.109.043321}, though the energies considered here receive contributions from both the electromagnetic and matter fields.  The functional forms for the energies in Eqs.~\eqref{eq:energies} manifest their respective signs. Assuming \(\psi_+\) has zero nodes, \(\psi_-\) has one, and their energies increase with the number of nodes, then the signs of all parameters are determined \(C_\text{J},d,E_\text{J1}>0\) and \(E_{\text{J}2}<0\), which in the absence of an external electric field, determines the ground state of Eq.~\eqref{eq:hamiltonian}:
\begin{equation}\label{eq:ground}
    (n_*,\phi_*) =
    \begin{cases}
        \left(0,0\right) & -E_{\text{J}1} < 4 E_{\text{J}2}\\
        \left(0,\pm\arccos{\left(-\frac{E_{\text{J}1}}{4E_{\text{J}2}}\right)}\right) & -E_{\text{J}1} > 4 E_{\text{J}2}.
    \end{cases}
\end{equation}
A derivation of the functional forms in Eqs.~\eqref{eq:energies}-\eqref{eq:parity_energies}, along with a proof of Eq.~\eqref{eq:ground}, is given in Appendix~\ref{sec:ground}. For all separation lengths \(L\) separating the large islands \(a \gg \xi\) considered in Fig.~\ref{fig}, the numerically obtained values of \(E_{\text{J}1,2}\) correspond to the trivial ground state. The doubly degenerate non-trivial ground state, known in the literature as the \(\phi\) JJ, is a leading candidate for on-chip phase batteries and has been experimentally realized in superconductor-ferromagnet-superconductor (SFS) junctions~\cite{PhysRevLett.109.107002, strambini2020josephson}. It is not surprising that we find a trivial ground state for the simple geometry considered here, though further study is needed to determine if a \(\phi\) JJ can be produced from more complex geometries. 

We proceed by analyzing the Hamiltonian in the limit of a small charge imbalance by expanding \(H\) up to second order in \(Q\)
\begin{widetext}
\begin{subequations}
    \begin{gather}
    H \approx H^{(0)} + H^{(1)} + H^{(2)}\\
    H^{(0)} = -E_\text{J1}\cos\left(2\pi\Phi/\Phi_0\right)-E_\text{J2}\cos\left(4\pi\Phi/\Phi_0\right)\\
    H^{(1)} = -dV'_\text{ext}Q\\
    H^{(2)} = \left(\frac{1}{2C_\text{J}} + \frac{4E_{\text{J}2}}{Q_0^2}+ \frac{2E_{\text{J}1}}{Q_0^2}\cos\left(2\pi\Phi/\Phi_0\right) + \frac{4E_{\text{J}2}}{Q_0^2}\cos\left(4\pi\Phi/\Phi_0\right)\right)Q^2
    \end{gather}
\end{subequations}
\end{widetext}
The EOM arising from \(H^{(0)}\) is the first Josephson relation with a second harmonic CPR
\begin{equation}\label{eq:2cpr_josephson}
    I \equiv -\dot{Q} = \underbrace{\frac{2\pi E_{\text{J}1}}{\Phi_0}}_{I_{\text{c}1}}\sin\left(2\pi\Phi/\Phi_0\right)+\underbrace{\frac{4\pi E_{\text{J}2}}{\Phi_0}}_{I_{\text{c}2}}\sin\left(4\pi\Phi/\Phi_0\right).
\end{equation}
As shown in Fig.~\ref{fig}(e), for large islands \(a\gg \xi\), the magnitudes of both critical currents \(I_{\text{c}1,2}\) decay with increasing separation length \(L\), and for \(L\gg\xi\), the first term in Eq.~\eqref{eq:2cpr_josephson} dominates the second \((\abs{I_{\text{c}2}/I_{\text{c}1}}\ll 1\)), thus reproducing the celebrated sinusoidal CPR. Inclusion of \(H^{(1)}\) produces the second Josephson relation
\begin{equation}
    \dot{\Phi} = -dV'_\text{ext} \equiv \Delta V
\end{equation}
where \(\Delta V\) is the voltage \emph{drop} measured from an electrode positioned at \(x=-d/2\) to one positioned at \(x=d/2\), a consequence of Eq.~\eqref{eq:potential} with \(\abs{\tilde{\alpha}_\text{L}}^2\approx\abs{\tilde{\alpha}_\text{R}}^2\). We recover the traditional interpretation of the second Josephson relation by noting that for large islands \(a\gg \xi\), we may approximate \(\psi_\pm(x>0) \approx \sqrt{\rho/q} \implies d \approx L+a\) [Eqs.~\eqref{eq:background},~\eqref{eq:d}] to identify \(\Delta V\) (the voltage drop from the left island center to the right) as the voltage drop across the junction. With the same approximation, the capacitance reduces to \(C_\text{J}\approx \epsilon_0 \sigma/(L+3a/2)\) [Eqs.~\eqref{eq:background},~\eqref{eq:C_J}], which for sufficiently large \(a\) and/or \(L\), causes the first term in \(H^{(2)}\) to dominate the rest. To second order in \(Q\), the Hamiltonian is therefore readily quantized
\begin{equation}\label{eq:charge_qubit}
    \hat{H} \approx \frac{\hat{Q}^2}{2C_\text{J}} + \hat{Q}\Delta V -E_{\text{J}1}\cos\left(2\pi\hat{\Phi}/\Phi_0\right) -E_{\text{J}2}\cos\left(4\pi\hat{\Phi}/\Phi_0\right)
\end{equation}
with the canonical commutation relation \([\hat{\Phi},\hat{Q}]=i\hbar\). We note that for \(L\gg a\), the capacitance reduces to that of two parallel plates separated by a distance \(L\), as expected, and for \(L\gg \xi\), the final term in Eq.~\eqref{eq:charge_qubit} may be dropped, thus producing the charge qubit Hamiltonian and laying the foundation for cQED.

To summarize the results of this study, we have provided an \emph{ab initio} derivation of the Josephson relations from the theory of non-relativistic scalar electrodynamics, a correction to the canonical CPR in the form of a second harmonic term, and a derivation of the charge qubit Hamiltonian that relates traditionally free qubit parameters to the junction's geometric and material properties. Future theoretical work should consider a gauge-invariant formulation of the analysis presented here, the study of more complex geometries (which may produce a non-trivial ground state), and a numerical solution of the full nonlinear partial differential equations [Eqs.~\eqref{eq:EOM_simple}] to establish the regime of validity of the two-mode approximation [Eq.~\eqref{eq:ansatz}]. We expect this approximation to be valid in the large separation limit, where the stationary parity states are nearly degenerate. \red{Finally, while the 1D treatment considered here reproduces many of the Josephson junction’s well-known physical properties (e.g., the Josephson relations), microscopic derivations predict that a spatially varying phase difference between extended junctions should obey the sine-Gordon equation~\cite{PhysRevB.45.12457}. An analogous derivation from the present theory would require relaxing the assumption of translational symmetry in one direction, which we leave to future work. 
}

\begin{acknowledgments}
The authors are especially grateful to Dung Pham, Zoe Zager, and Wentao Fan for many insightful discussions. This material is based upon work supported by the National Science Foundation Graduate Research Fellowship under Grant No. DGE-2039656 and by the US Department of Energy,
Office of Basic Energy Sciences, Division of Materials Sciences and Engineering, under Award No. DESC0016011.
\end{acknowledgments}

\appendix
\section{Numerics}\label{sec:numerics}
    In terms of the unitless variables
    \begin{subequations}
        \begin{gather}
            \bar{\psi}_\pm \equiv \sqrt{\frac{\mu_0}{m}}\abs{q}\lambda_\text{L}\psi_\pm\\
            \bar{V}_\pm \equiv \frac{q\lambda_\text{L}}{\hbar c}V_\pm\\
            \bar{\grad} \equiv \xi \grad\\
            \bar{\mathbf{x}} \equiv \frac{\mathbf{x}}{\xi}\\
            \bar{L} = \frac{L}{\xi},
        \end{gather}
    \end{subequations}
    the equations of state (Eqs.~\eqref{eq:EOM_simple} with \(\dot{\psi}_\pm=0\)) read
    \begin{subequations}
        \begin{gather}
            \left(-\bar{\nabla}^2 + \bar{V}_\pm\right)\bar{\psi}_\pm=0\\
            2\bar{\nabla}^2\bar{V}_\pm + \abs{\bar{\psi}_\pm}^2 = \Theta\left(\abs{\bar{x}}-\bar{L}/2\right).\label{eq:nondim_gauss}
        \end{gather}
    \end{subequations}
    with the boundary conditions 
    \begin{subequations}
        \begin{gather}
            \bar{\psi}_\pm(-\bar{x})=\pm\bar{\psi}_\pm(\bar{x})\\
            \bar{V}_\pm(-\bar{x}) = \bar{V}_\pm(\bar{x})\\
            \lim_{\bar{x}\rightarrow\infty}\bar{\psi}_\pm(\bar{x})=1\\
            \lim_{\bar{x}\rightarrow\infty}\bar{V}_\pm(\bar{x})=0.
            \end{gather}
    \end{subequations}
    We solve these equations for 21 equally spaced separation lengths spanning the range \(\bar{L} \in [0,40]\) on a grid of size \(\bar{x} \in [0,100]\) with spacing \(d\bar{x}=0.25\). Using a central finite difference scheme with second order accuracy for the derivatives, we convert the differential equations into a set of algebraic ones and find their numerical solution via Netwon's method. All solutions converged to machine precision (10 decimal places) using the \texttt{findroot} function from the \texttt{mpmath} library in Python with the initial guess \(\bar{\psi}_\pm = 1\) and \(\bar{V}_\pm = 0\). Finally, we compute the critical currents using the rectangle rule for the integrals
    \begin{subequations}
        \begin{gather}
            I_{\text{c}1} = -\frac{1}{4}\left[\frac{mc \xi \sigma}{\mu_0 q \lambda_\text{L}^3}\right]\int_{-\infty}^\infty \left(\abs{\bar{\psi}_-}^2+\abs{\bar{\psi}_+}^2\right)\left(\bar{V}_--\bar{V}_+\right)d\bar{x}\\
            I_{\text{c}2} = -\frac{1}{8}\left[\frac{mc \xi \sigma}{\mu_0 q \lambda_\text{L}^3}\right]\int_{-\infty}^\infty \left(\abs{\bar{\psi}_-}^2-\abs{\bar{\psi}_+}^2\right)\left(\bar{V}_--\bar{V}_+\right)d\bar{x}.
        \end{gather}
    \end{subequations}
\section{Two-mode approximation}\label{sec:2mode}
    We derive the EOM in the two-mode approximation by first plugging the ansatz [Eq.~\eqref{eq:ansatz}] into Schrödinger's equation (Eq.~\eqref{eq:schrodinger}) and using Eq.~\eqref{eq:schrodinger} for \(\psi_\pm\) to eliminate the kinetic terms
    \begin{equation}
        \begin{aligned}
            i\hbar \sum_\pm\dot{\alpha}_\pm\psi_\pm &= \sum_\pm\left(-\frac{\hbar^2}{2m}\laplacian+qV\right)\alpha_\pm\psi_\pm\\
            &= q\sum_\pm\left(V-V_\pm\right)\alpha_\pm\psi_\pm.
        \end{aligned}
    \end{equation}
    We then integrate against \(\psi_\pm\), whose orthogonality \(\int_{-\infty}^\infty\psi_+\psi_-dx=0\) produces Eq.~\eqref{eq:2mode_parity}, and move to the rotating frame \(\tilde{\alpha}_{\pm} \equiv \alpha_{\pm} \exp{i\int _{-\infty}^t (\Delta_{++}+\Delta_{--})dt/(2\hbar)}\), where Schrödinger's equation reads
    \begin{equation}
        i\hbar\begin{pmatrix}\label{eq:two_paramter_schrodinger}
        \dot{\tilde{\alpha}}_+\\
        \dot{\tilde{\alpha}}_-
        \end{pmatrix}
         =
        \underbrace{\begin{pmatrix}
            K & U \\
            U & -K
        \end{pmatrix}}_{\mathcal{H}}
        \begin{pmatrix}
            \tilde{\alpha}_+\\
            \tilde{\alpha}_-
        \end{pmatrix}
    \end{equation}
with the matrix elements given by Eqs.~\eqref{eq:first_K}-\eqref{eq:first_U}. That \(\Delta_{+-}=\Delta_{-+}\) can be seen by dropping the odd contributions to the integrands for \(\Delta_{ij}\). The final step amounts to a simple basis change
    \begin{multline}
        i\hbar
        \begin{pmatrix}
            \dot{\tilde{\alpha}}_\text{R}\\\dot{\tilde{\alpha}}_\text{L}
        \end{pmatrix}
        = 
        \begin{pmatrix}\frac{1}{\sqrt{2}}&\frac{1}{\sqrt{2}}\\\frac{1}{\sqrt{2}}&-\frac{1}{\sqrt{2}}\end{pmatrix}
        \mathcal{H}
        \begin{pmatrix}\frac{1}{\sqrt{2}}&\frac{1}{\sqrt{2}}\\\frac{1}{\sqrt{2}}&-\frac{1}{\sqrt{2}}\end{pmatrix}
        \begin{pmatrix}
            \tilde{\alpha}_{\text{R}}\\\tilde{\alpha}_{\text{L}}
        \end{pmatrix},
    \end{multline}
    which after explicit matrix multiplication yields Eq.~\eqref{eq:2mode_just_dynamics}. 
\section{Electric potential}\label{sec:potential}
    We compute the electric potential by first eliminating the background charge density in favor of the stationary parity states as follows
    \begin{equation}
        \begin{aligned}
            &-\laplacian\left(V-\sum_\pm\abs{\tilde{\alpha}_\pm}^2V_\pm\right)\\ 
            &\quad\quad= \frac{1}{\epsilon_0}\left(q\abs{\psi}^2+cj_0\right)- \frac{1}{\epsilon_0}\sum_\pm\abs{\tilde{\alpha}_\pm}^2\left(q\abs{\psi_\pm}^2 + cj_0\right)\\
            &\quad\quad= \frac{q}{\epsilon_0}\left(\abs{\psi}^2-\sum_\pm\abs{\tilde{\alpha}_\pm}^2\abs{\psi_\pm}^2\right)\\
            &\quad\quad= \frac{2q}{\epsilon_0}\Re{\tilde{\alpha}_+^*\tilde{\alpha}_-}\psi_+\psi_-.
        \end{aligned}
    \end{equation}
    In the first equality, we have used Eq.~\eqref{eq:gauss}, in the second equality, we have used the fact that the probability amplitudes sum to unity \(\sum_\pm\abs{\tilde{\alpha}_\pm}^2=1\), and in the third equality, we have expanded \(\psi\) in the parity basis and canceled like terms. Up to an unphysical global phase, double integration yields
    \begin{multline}\label{eq:supp_V_1}
        V= -\frac{2q}{\epsilon_0}\Re{\tilde{\alpha}_+^*\tilde{\alpha}_-}\int_0^x \int_{0}^{x'}\prod_\pm\psi_\pm(x'')dx''dx'\\+f(t)x+\sum_\pm\abs{\tilde{\alpha}_\pm}^2V_\pm
    \end{multline}
    with \(f(t)\) the constant of integration, which is related to the boundary condition \(V'_\text{ext}(t)\) by
    \begin{equation}
        V'_\text{ext}(t) = f(t) - \frac{2q}{\epsilon_0}\Re{\tilde{\alpha}_+^*\tilde{\alpha}_-}\int_0^\infty \prod_\pm\psi_\pm(x'')dx''.
    \end{equation}
    Eliminating \(f(t)\) from Eq.~\eqref{eq:supp_V_1}  in favor of \(V'_\text{ext}(t)\) and rewriting the coefficients in the left/right basis produces Eq.~\eqref{eq:potential}. 
\section{Matrix elements}\label{sec:matrix_elements}
    We begin by noting that while Eq.~\eqref{eq:potential} is more compact than Eq.~\eqref{eq:supp_V_1}, it obfuscates the fact that the final term (the sum) is even, while all other terms are odd, a property that is manifest in Eq.~\eqref{eq:supp_V_1}. By dropping all odd contributions to the integrands for \(\Delta_{ij}\), we have
    \begin{widetext}
    \begin{subequations}
        \begin{gather}
           K = \frac{q}{2\mathcal{N}}\int_{-\infty}^\infty \left(\left(\sum_\pm\left(\frac{1}{2}\pm\Re{\tilde{\alpha}_\text{L}^*\tilde{\alpha}_\text{R}}\right)V_\pm - V_+\right)\abs{\psi_+}^2-\left(\sum_\pm\left(\frac{1}{2}\pm\Re{\tilde{\alpha}_\text{L}^*\tilde{\alpha}_\text{R}}\right)V_\pm - V_-\right)\abs{\psi_-}^2\right)dx
           \\
           U = \frac{q}{\mathcal{N}}\int_{-\infty}^\infty \left(\frac{q}{\epsilon_0}\left(\abs{\tilde{\alpha}_\text{R}}^2-\abs{\tilde{\alpha}_\text{L}}^2\right)\int_0^x \int_{x'}^{\infty} \prod_\pm\psi_\pm(x'')dx''dx'+V'_\text{ext}(t)x\right)\psi_+(x)\psi_-(x)dx.
        \end{gather}
    \end{subequations}
    \end{widetext}
    which after factoring out the terms that depend on the coefficients, yields Eqs.~\eqref{eq:K}-\eqref{eq:U}.

\section{Hamiltonian formulation}\label{sec:hamiltonian}
In terms of the Madelung variables, the EOM read
\begin{subequations}
    \begin{gather}
        \dot{n}_R = -\frac{2K\sqrt{n_L n_R}}{\hbar}\sin\phi\label{eq:nR}\\
        \dot{n}_L = \frac{2K\sqrt{n_L n_R}}{\hbar}\sin\phi\label{eq:nL}\\
        \dot{\phi}_R = \frac{1}{\hbar}\left( -K\sqrt{\frac{n_L}{n_R}}\cos\phi-U\right)\label{eq:phiR}\\
        \dot{\phi}_L = \frac{1}{\hbar}\left(-K\sqrt{\frac{n_R}{n_L}}\cos\phi+ U\right)\label{eq:phiL}.
    \end{gather}
\end{subequations}
Subtracting Eqs.~\eqref{eq:nR} and \eqref{eq:phiL} from Eqs.~\eqref{eq:nL} and \eqref{eq:phiR}, respectively, yields Eqs.~\eqref{eq:n_phi_EOM}, which may equivalently be stated as EOM for the charge imbalance 
\begin{multline}
    \dot{Q} = -\frac{2\pi E_{J1}}{\Phi_0}\left(1-\left({2Q}/{Q_0}\right)^2\right)^{1/2}\sin\left(2\pi\Phi/\Phi_0\right)
         \\-\frac{8\pi E_{J2}}{\Phi_0}\left(1-\left(2Q/Q_0\right)^2\right)\sin\left(2\pi\Phi/\Phi_0\right)\cos\left(2\pi\Phi/\Phi_0\right)
\end{multline}
and the flux
\begin{multline}
    \dot{\Phi} = -dV'_\text{ext} + \frac{Q}{C_\text{J}}
        \\+\frac{2E_{J1}}{Q_0}\left(2Q/Q_0\right) \left(1-\left(2Q/Q_0\right)^2\right)^{-1/2}\cos\left({2\pi\Phi/\Phi_0}\right) \\+ {\frac{8E_{J2}}{Q_0}\left(2Q/Q_0\right)\cos^2\left({2\pi\Phi/\Phi_0}\right)}.
\end{multline}
Explicit calculation shows these are Hamilton's equations \(\dot{\Phi} = \partial_Q H\) and \(\dot{Q} = -\partial_\Phi H\) with \(H\) the Hamiltonian in Eq.~\eqref{eq:hamiltonian}.
\section{Ground state}\label{sec:ground}
    We begin by casting \(E_{\text{J}1,2}\) in a form that makes their signs clear. Alternating application of Eqs.~\eqref{eq:EOM_simple} and integration by parts yields
\begin{equation}
    \begin{aligned}
        &E_{\text{J1}}\\
        &= \sum_\pm\frac{\sigma}{4}\int_{-\infty}^\infty\left(\epsilon_0\laplacian V_\pm +cj_0\right)\left(V_--V_+\right)dx\\
        &= \sum_{\pm}\pm\frac{\sigma}{2}\int_{-\infty}^\infty\left(\frac{\epsilon_0}{2} \abs{\grad V_\pm}^2 - cj_0V_\pm\right)dx\\
        &= 
        \sum_\pm \pm \frac{\sigma}{2}\int_{-\infty}^\infty\left(\frac{\epsilon_0}{2} \abs{\grad V_\pm}^2 +\left(q\abs{\psi_\pm}^2 + \epsilon_0 \laplacian V_\pm\right)V_\pm\right)dx \\
        &= 
        \sum_\pm \mp\frac{\sigma}{2}\int_{-\infty}^\infty\left(\frac{\epsilon_0}{2} \abs{\grad V_\pm}^2 -q\abs{\psi_\pm}^2V_\pm\right)dx \\
        &= 
        \sum_\pm \mp \frac{\sigma}{2}\int_{-\infty}^\infty\left(\frac{\epsilon_0}{2} \abs{\grad V_\pm}^2 -\frac{\hbar^2}{2m}\psi_\pm \laplacian \psi_\pm\right)dx \\
        &= 
        \sum_\pm \mp\frac{\sigma}{2}\int_{-\infty}^\infty\left(\frac{\epsilon_0}{2} \abs{\grad V_\pm}^2 +\frac{\hbar^2}{2m}\abs{\grad\psi_\pm}^2\right)dx \\
        &= \frac{E_--E_+}{2},
    \end{aligned}
\end{equation}
which is positive assuming the antisymmetric stationary parity state has higher energy than the symmetric one, and
\begin{equation}
    \begin{aligned}
        &E_{\text{J}2}\\&= -\frac{\sigma}{16}\int_{-\infty}^\infty\left(-\epsilon_0\laplacian V_- + \epsilon_0 \laplacian V_+ \right)\left(V_--V_+\right)dx\\
        &= -\frac{\epsilon_0\sigma}{16}\int_{-\infty}^\infty\abs{\grad\left(V_--V_+\right)}^2dx,
    \end{aligned}
\end{equation}
which is necessarily negative. The assumption that \(\psi_+\) has no nodes and \(\psi_-\) has one (at the origin) implies \(\psi_\pm(x>0)>0\). Since the integrals in Eqs.~\eqref{eq:C_J}-\eqref{eq:d} depend only on the stationary parity states evaluated at positive spatial coordinates, the integrands (and therefore integrals) are strictly positive, yielding \(C_\text{J}, d >0\). We seek the global minimum of the Hamiltonian \((n_*,\phi_*)\) in the absence of an external electric field, which will coincide with the global minimum of 
\begin{equation}
    \bar{H}(n,\phi) \equiv n^2 -\bar{E}_{\text{J}1}\sqrt{1-n^2}\cos\phi -2\bar{E}_{\text{J}2}\left(1-n^2\right)\cos^2\phi,
\end{equation}
where the scaled Josephson energies \(\bar{E}_{\text{J}1,2}\equiv (2C_\text{J}/Q_0^2)E_{\text{J}1,2}\) have the same signs as the original ones \(E_{\text{J}1,2}\). The global minimum must satisfy \(\partial_\phi \bar{H}(n_*,\phi_*)= 0\), which implies
\begin{equation}\label{eq:conditions}
    n_* = 1~\lor~\sin\phi_*=0~\lor~-\frac{\bar{E}_\text{J1}}{4E_\text{J2}} = \sqrt{1-n_*^2}\cos\phi_*
\end{equation}
The first condition in Eq.~\eqref{eq:conditions} cannot be true, since \(\bar{H}(0,\pi/2)<\bar{H}(1,\phi)~\forall\phi\). We must therefore have \(n_*\neq 1\). The second condition in Eq.~\eqref{eq:conditions} implies \(\phi_* = 0~\lor~\phi_* = \pi\), but \(\partial^2_\phi \bar{H}(n,\pi) < 0~\forall n\), so \(\phi_* \neq \pi\). The remaining conditions thus read
\begin{equation}
    \phi_* \in \left\{0,\pm \arccos\left(-\frac{\bar{E}_{\text{J}1}}{4\bar{E}_{\text{J}2}\sqrt{1-n_*^2}}\right)\right\}.
\end{equation}
Noting that \(\partial_n \bar{H}(n,0) = 0 \implies \partial_\phi^2\bar{H}(n,0) < 0\), we must have \(\phi_* = 0 \implies n_* = 0\). Moreover, 
\begin{equation}
    \bar{H}\left(n, \pm \arccos\left(-\frac{\bar{E}_{\text{J}1}}{4\bar{E}_{\text{J}2}\sqrt{1-n^2}}\right)\right) = \frac{\bar{E}^2_{\text{J}1}}{8 \bar{E}_{\text{J}2}} + n^2
\end{equation}
is clearly minimized for \(n = 0\), so \(n_* = 0\) for all possible ground states. The two possible ground state energies are \(\bar{H}(0,0) = -\bar{E}_{\text{J}1}-2\bar{E}_{\text{J}2}\) and \(\bar{H}(0,\pm\arccos(-\bar{E}_{\text{J}1}/(4\bar{E}_{\text{J}2}))=\bar{E}_{\text{J}1}^2/(8\bar{E}_{\text{J}2})\). The minimum of these two energies depends on the ratio of Josephson energies as described by Eq.~\eqref{eq:ground}.

\bibliography{refs}

\end{document}